\documentclass[runningheads]{llncs}

\usepackage[T1]{fontenc}

\usepackage{graphicx}
\usepackage{hyperref}

\hypersetup{breaklinks=true} 
\usepackage{url}             

\usepackage[a-1b]{pdfx}
\usepackage{amsmath}
\DeclareFontFamily{U}{skulls}{}
\DeclareFontShape{U}{skulls}{m}{n}{ <-> skull }{}

\mathchardef\UrlBreakPenalty=1000
\mathchardef\UrlBigBreakPenalty=1000

\begin{document}

\title{Lessons from a Big-Bang Integration: Challenges in Edge Computing and Machine Learning}

\titlerunning{Lessons from a Big-Bang Integration}

\author{Alessandro Aneggi\orcidID{0009-0009-3709-5051},Andrea Janes\orcidID{0000-0002-1423-6773}}

\authorrunning{Aneggi, Janes}

\institute{Free University of Bozen-Bolzano, Bolzano, Italy \\
\email{\{aaneggi, ajanes\}@unibz.it}}
\maketitle              
\begin{abstract}
This experience report analyses a one-year project focused on building a distributed real-time analytics system using edge computing and machine learning. The project faced critical setbacks due to a “big-bang” integration approach, where all components—developed by multiple geographically dispersed partners—were merged at the final stage. The integration effort resulted in only six minutes of system functionality, far below the expected 40 minutes. Through root cause analysis, the study identifies technical and organisational barriers, including poor communication, lack of early integration testing, and resistance to top-down planning. It also considers psychological factors such as a bias toward fully developed components over mock-ups. The paper advocates for early mock-based deployment, robust communication infrastructures, and the adoption of top-down thinking to manage complexity and reduce risk in reactive, distributed projects. These findings underscore the limitations of traditional Agile methods in such contexts and propose simulation-driven engineering and structured integration cycles as key enablers for future success.

\keywords{big-bang integration \and reactive applications \and retrospective}

\end{abstract}
\section{Introduction}

This report reflects on last year's project, where a ``Big-Bang Integration'' caused a missed deadline by merging all components at once instead of gradually. Unfortunately, since ``developers struggle to know if what they are doing makes sense for anyone else'' \cite{Mellegard2020}, such a single integration step often leads to unexpected severe consequences.

\begin{figure}[ht]
    \centering
    \includegraphics[width=.6\linewidth]{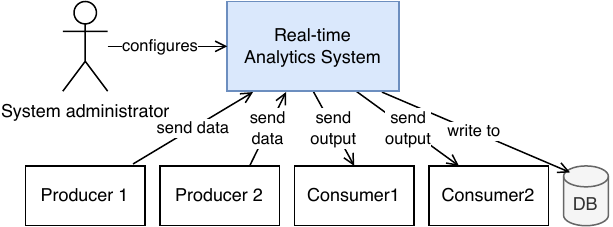}
    \caption{Architecture (using the notation of a C4-model context diagram \cite{Brown2023})}
    \label{fig:architecture}
\end{figure}

The project consists of developing a real-time analytics system using edge computing, where researchers and industry professionals work as distributed teams on different components. An architectural overview of the created system is depicted in Fig. \ref{fig:architecture}: two producers of data form the input of a real-time system, which analyses the data and produces the output for two consumers. They were developed by five different project partners.

The described system poses significant challenges as it is a reactive system \cite{reactive2014}, processing sensor data in real time and forwarding the output to consumers. The database shown in Fig.~\ref{fig:architecture} acts as a backup for incoming data in case the analytics subsystem fails.
Producers are software components that generate data from sensors or other sources. Consumers, on the other hand, are components that use data generated either internally or from external sources.
From the client’s perspective, such systems are binary: they either work or they don't. For example, if the system processes the data correctly but fails to deliver the output to consumer two, it is still perceived as a failure.

The integration took place in two days at a location were the system could be tested. The idea was to do a first test one the day before the deadline and to make the final adjustments so that at the deadline the final product could be demonstrated. Already during the first day, we encountered several issues, e.g.: some hardware was initially not present and had to be retrieved; the network on the testing site was much slower than expected, this required the team to modify the source code on the first day to send data with a lower resolution; components had performance issues not encountered when developing them at the company sites; an important team member did not attend the integration, which transform in spend time to solve som issue.

\section{Observations and Insights}
\label{sec:observations}

Following the GQM approach, we formulated the following questions: 

Q1. \textit{What were the issues with the integration?} 

Q2. \textit{Why did the team decide to build the software using a bottom-up approach?} 

Q3. \textit{Were there any early warning signs indicating that the deadline might not be met?} 

We observed various challenges (answer to Q1): 
 
1) research projects often require not only time for development but also time to discover what is yet not unknown; 

2)  collaboration across geographically dispersed teams from different organisations made it complex to expose problems;

    3)  it was complicated and expensive to test the entire system with all its parts, since expensive equipment and distributed team;
    
    4)  only team leads had weekly meetings, sometimes resulting in ``Chinese whispers'' among researchers; 
    
    5)  developing systems in real-time processing, particularly within the framework of a reactive application, require the attention to details.
Considering the fact that all resources are limited, the question is how to achieve the project goals with the given resources minimising the risk.

Agile methodologies often promote iterative development and early integration \cite{Pasuksmit2024}.

We understand this as promoting a ``top-down'' development approach in which one begins deploying a system that instead of real, functioning components just contains mock components, i.e., components that imitate the final ones. This would mean that the first deployed system should have been as the one in Fig. \ref{fig:architecture}, where producers and consumers are either doing nothing or producing test data.

As components are developed, mock components are replaced with implemented counterparts, which however, is not a ``minimal viable product'' (MVP) as nothing functions. It represents just the deployment of the architecture, so that one can test newly developed features immediately in a production-like environment.

Studying why the team chose to pursue a bottom-up approach (answer to Q2), we observed that producing test data was extremely difficult.
Let's assume, one wants to build a face recognition door lock system, which cannot be circumvented using a photograph, one that uses 3D cameras. To test such a system I need to test it with real humans or obtain or create a testing dataset and use that as a testing input for the developed door lock system. 
In our case, such a data set did not exist, could not be easily created and manual testing was expensive. 
This leads to the first insight: \textbf{In projects where integration testing is difficult, we will dedicate time and foresee it in the project plan to create testing data to setup a mock architecture as soon as possible.}

In many projects involving machine learning and AI (as the present one), the development of the individual components can be very complex and require specialized researchers in particular areas. 
This leads to the second insight: \textbf{in projects where distributed components must work together to deliver the project's output, we will dedicate as much attention and resources to the communication infrastructure as we do to the components themselves.}

To further analyze Q2, 

we conducted a web search on Google on December 16th, 2024. The search was performed in incognito mode using the query ``software engineering top-up top-down'', and we reviewed the results from the first two pages. 
From this analysis Top-down approaches focus on planning and system-wide understanding \cite{lit4} before implementation but can be inflexible and require more upfront resources \cite{lit8}. They involve less communication between modules \cite{lit3} and may struggle with unknowns at the start \cite{lit3}\cite{lit4}. 
Bottom-up approaches prioritize coding and early testing, making them ideal for rapid prototyping\cite{lit5}\cite{lit7}. However, they risk poor integration if modules are developed without a clear linking strategy. Bottom-up is favored for its focus on fundamentals before addressing the whole system \cite{lit6}.

These results suggest practitioners hesitate to view the system as a whole, which is key for early integration. The statements analysed above imply that top-down thinking requires full system understanding or works only with minimal module communication. We strongly disagree with these views.

This brief web search aligned with our project experience: the belief that the system could not yet be fully defined led to a bottom-up approach. There might also be psychological factors playing a role, including: a) it seems more productive to ``fully develop'' components rather than creating mock components, and b) it is more rewarding to build something that works than to create a component that is incomplete but uncovers communication issues with other components. This last point, uncovering communication problems, requires an open error culture within the team; otherwise, the issue is merely attributed to the component being ``incomplete.'' 

This brings us to the third insight: \textbf{in projects where distributed components must work together to deliver the project's output, we have to establish a top-down mindset within the team so that all know why it is worth to a) investing in building mock components and b) performing integration testing as early as possible.}

To answer Q3 (if there were early warning signs), we list the warning signs we observed during the project: 

   \textbf{ 1) Procrastination in decision-making}: Delays in reaching agreements on standards or resolving specific issues often led to bottlenecks. E.g. critical decisions were postponed until late in the project timeline.
   
    \textbf{ 2) Overemphasis on individual solutions:}  Many participants focused predominantly on optimising their own components rather than considering the system as an integrated whole.
    
    \textbf{ 3) Lack of shared prototypes for testing:} Teams often failed to provide others with working prototypes or sample data necessary for testing. 
    
    \textbf{ 4) Insufficient feedback mechanisms:}  The absence of robust communication and feedback systems between partners led to misalignment.
    
    \textbf{ 5) Operating too close to technological limits:}  The system frequently operated near the threshold. E.g. bandwidth utilization of 9.6 GB out of a 10 GB limit and server CPUs running at over 90\% capacity.
    
    \textbf{ 6) Last-minute changes after integrated testing:} As a consequence of the last-day integrated test, some parameters were changed overnight, forcing other partners to adapt without sufficient time for testing.
\section{Conclusions}
\label{sec:conclusion}
This experience report described our observations and insights of the development of a distributed reactive real-time system involving hardware and software. 

Our key observation is that, although Agile methodologies promote early integration and continuous testing, their effective implementation remains challenging in multi-partner, distributed projects. 
For instance, while collaboration tools and regular cross-team meetings seem to solve distributed team challenges easily, in projects like ours with multiple partners, each has established ways of working and may resist changing their approach solely for the sake of one project. 
Similarly, while continuous integration and automated testing address minimal early testing, in our case, some teams implemented these practices independently without a unified, coordinated plan.

\textbf{We interpreted those findings using root cause analysis.} Beyond technical difficulties, the primary drivers of the integration challenges were psychological factors such as a preference for 'finished' components over mock-ups, a lack of shared ownership of system-level risks, and communication barriers across organisational boundaries.

We want to prioritize the lessions learned as follows: first, unified communication infrastructure must be prioritized from the start, on par with technical development; second mock-based early deployment of the system architecture is critical, even if real components are not ready, and third, top-down thinking (intended as seeing the system as a whole rather than isolated parts) must be actively promoted within teams. Finally a better understanding and awareness of software management techniques specifically tailored for reactive projects can help reduce development risks. 

\textbf{Proposed Practices for Future Projects.} To address these issues, we propose multiple action as: establish a shared simulation-driven integration environment from project inception, ---even with only mock components---; one team as responsible of cross-integration demos and explicit resources and planning time to the creation of realistic test data and mock environments early in the project.
Moreover, we suggest that adopting practices such as simulation-driven engineering \cite{Meyers2024} can lead to better outcomes and faster delivery timelines. When possible, managers and developers should favor the use of effective tools and development techniques over simply adding more meetings or resources, as this can lead to more efficient and successful project execution.

\textbf{Open Questions for Further Research.}

  1) How can effective collaboration be ensured between teams with differing goals? While coordination within a company or between aligned partners is straightforward, collaboration becomes more difficult when partners have diverging interests or incentives.

 2) Can integration testing mitigate the effects of weak organisational alignment? Even if not all stakeholders share the same goals, can a robust integration testing strategy ensure coherence and project success?

 3) Can specific tools or frameworks better support these projects? Would shifting from high-level planning to practical tooling help managers and developers stay goal-oriented and reduce unnecessary methodological overhead?

 4) What is the optimal trade-off between planning and execution? Meetings and formal coordination consume resources. How much planning is truly necessary to avoid inefficiency without losing critical alignment?

These open questions point toward a broader research agenda: evaluating how structured practices—such as mandatory cross-partner mock-based integration, simulation-driven environments, and lightweight tooling—can systematically improve coordination and resilience in reactive, distributed systems. Empirical studies or controlled experiments comparing these practices across similar project contexts could provide deeper insights and lead to evidence-based guidelines.

\textbf{Implications for Future Development Methodologies}
We suggest that current software development practices—particularly Agile—should be adapted to better support the unique challenges of reactive, distributed systems, especially in multi-partner environments. Reactive systems and collaborative projects introduce complexities that often exceed the capabilities of traditional project management techniques, increasing the risk of failure, particularly during large-scale integrations. Enhancing awareness of software management strategies tailored for reactive development, incorporating warning signals to prompt timely adjustments, and adopting approaches such as simulation-driven engineering \cite{Meyers2024} and lightweight, mandatory cross-partner integration cycles may reduce development risks, improve delivery timelines, and lead to more successful project outcomes.

\bibliographystyle{splncs04}
\bibliography{bibliography}

\end{document}